\documentclass[
twocolumn,
aps,prd,
nofootinbib,
superscriptaddress,
showpacs,
tightenlines,
amsmath,
amssymb
]{revtex4}
\usepackage{bm}
\usepackage[dvips]{color,graphicx}

\begin{document}
                               

\title{Microscopic Description of Deeply Virtual Compton Scattering off Spin-0 
Nuclei.}

\author{S.~Liuti}
\email[]{sl4y@virginia.edu}

\author{S.~K.~Taneja}
\email[]{skt6c@virginia.edu}

\affiliation{University of Virginia, Charlottesville, Virginia 22901, USA.}


\begin{abstract}
We evaluate within a microscopic calculation 
the contributions of both coherent and incoherent
deeply virtual Compton scattering from a spin-0 nucleus.  
The coherent contribution is obtained when the target nucleus recoils
as a whole, whereas for incoherent scattering 
break-up configurations for the final nucleus into a an outgoing nucleon
and an $A-1$ system are considered. 
The two processes encode different characteristics of generalized parton 
distributions.
\end{abstract}

\maketitle

Recent theoretical and experimental developments have identified a new frontier for
studies of the quark and gluon content of hadronic systems in a class of 
high energy exclusive processes. Their prototype is Deeply Virtual 
Compton Scattering (DVCS)  \cite{Diehl_hab}. 
DVCS provides 
observables that are most straightforwardly related to Generalized Parton Distributions
(GPDs), the new theoretical tools introduced in \cite{DMul1,Ji1,Rad1}
to describe in a partonic language 
the orbital angular momentum carried by the nucleon's 
constituents. GPDs were found more recently to describe also
partonic structure in the transverse direction with 
respect to the large longitudinal momentum in the reaction. 
They are, in fact, the Fourier
transforms of the so-called 
Impact Parameter dependent Parton Distributions Functions (IPPDFs) \cite{Bur}.
       
\begin{figure}
\includegraphics[width=7.5cm]{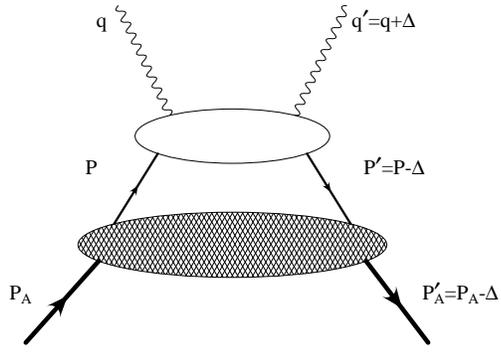}
\includegraphics[width=6.cm]{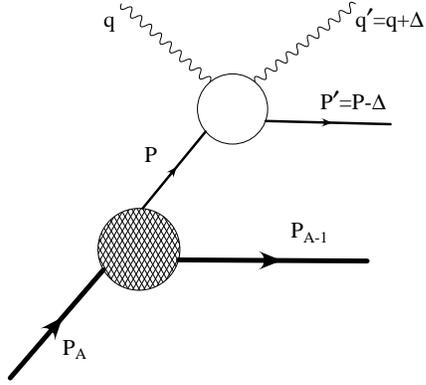}
\includegraphics[width=7.5cm]{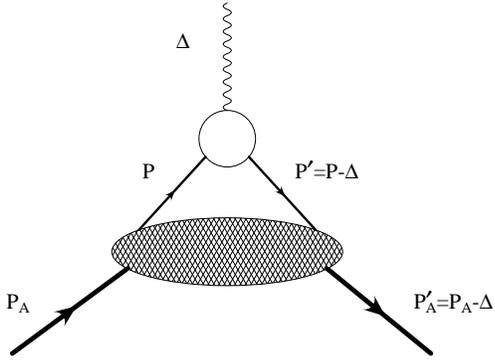}
\includegraphics[width=6.0cm]{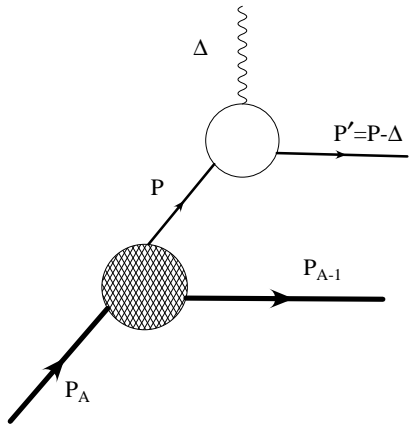}
\caption{Amplitudes for DVCS and BH processes from a nuclear target 
at leading order in $Q$. {\bf (a)} DVCS, coeherent process; {\bf (b)} DVCS, incoherent
process; {\bf (c)} BH, coeherent process; {\bf (d)} BH, incoherent process.} 
\label{fig1}
\end{figure}

DVCS on a proton target, $ep \rightarrow e^\prime p \gamma$, allows one to unravel 
in principle
four GPDs: the unpolarized quark distributions, $H^q(X,\zeta,t)$, 
and $E^q(X,\zeta,t)$, and
the polarized ones, $\widetilde{H}^q(X,\zeta,t)$, and $\widetilde{E}^q(X,\zeta,t)$.
The kinematical variables are: $P^\prime=P-\Delta$,  
$k^\prime=k-\Delta$, the final nucleon's and quark's momenta,
respectively; $q$, the virtual photon momentum, and $q^\prime = q+\Delta$,
the outgoing photon momentum. Furthermore: $-\Delta^2 = t$, 
$X=k^+/P^+$, $\zeta=\Delta^+/P^+$ (Fig.\ref{fig1}).
\footnote{
We use the notation: 
$ \displaystyle p^\pm = \frac{1}{\sqrt{2}}(p_o \pm p_3)$, with
$\displaystyle (pk) = p^+k^- + p^-k^+  - p_\perp \cdot k_\perp$. }  
GPDs, or the correlation functions in the diagram, correspond to 
``hybrid'' distributions in that they encode properties of both
Parton Distributions Functions (PDFs) from inclusive Deep Inelastic Scattering 
(DIS), and of the proton
form factors. In the (forward) limit: $t,\zeta \rightarrow 0$, $H^q(X,\zeta,t)=q(x)$,
and $\widetilde{H}^q(X,\zeta,t) = \Delta q(x)$, the PDFs for polarized and unpolarized
DIS, respectively. At the same time, GPDs first moments in $X$ are given by the proton
Dirac and Pauli form factors in the unpolarized case, and by the axial-vector and 
pseudo-scalar form factors in polarized scattering.  

The first DVCS measurements were performed at HERA \cite{HERMES,H1,ZEUS} and at 
CLAS \cite{CLAS}. Because of the presence of the Bethe-Heitler (BH) process producing
the same $e p \gamma$ final state and dominating the cross sections at currently
available kinematics, GPDs can be obtained only through the interference term between
the DVCS and BH processes. 
In Born approximation, one writes:
\begin{eqnarray}
T^{2}= |T_{DVCS}|^2 + |T_{BH}|^2 + \mathcal{I}
\label{T}
\end{eqnarray}
Where $\mathcal{I}$, the interference term, is given by,
\begin{eqnarray}
\mathcal{I} = \left[ T_{DVCS}T^{*}_{BH}+ T^{*}_{DVCS}T_{BH} \right]
\end{eqnarray} 
At leading order in the four-momentum transfer 
$Q \equiv (-q^2)^{1/2}$ (twist-two), the contribution of ${\cal I}$ to the
cross section is expressed as a finite 
sum of Fourier terms \cite{KirMul}:
\begin{eqnarray}
{\cal I} & = & 
{\cal K} \left[c_o^{\cal I} + \sum_{n=1}^3 \left(c_n^{\cal I} \cos (n\phi) + 
s_n^{\cal I} \sin (n \phi) \right) \right],  
\label{I_fourier}
\end{eqnarray}
where ${\cal K}$ is a kinematical factor, and $\phi$ is the azimuthal angle between 
the lepton plane and the scattered hadron plane. 
The dominant harmonics are for $n=1$, the higher order ones being suppressed by 
$\alpha_S$ \cite{KirMul}. 
Several asymmetries (beam spin, $A_{LU}$, charge, $A_C$, target spin $A_{UL}$, etc...) 
have been identified,  through which one can access ${\cal I}$, and probe
different GPDs, or components of GPDs 
(see {\it e.g.} \cite{Diehl_hab}).

GPDs were recently also measured through lepto-production of a photon 
off nuclear targets, $eA \rightarrow e^\prime \gamma A$ 
\cite{Hasch}. The study of nuclear targets is particularly important as they 
provide a laboratory where additional information
can be obtained on these elusive observables. Exploratory studies 
using GPDs were performed both on the phenomenon of Color Transparency
\cite{PirRal_DVCS,BurMil,LiuTan1}, {\it i.e.} on the rate of 
survival of small size hadronic configurations as 
the quasi-elastically struck nucleons scatter through the nuclear medium, 
and on the ``generalized'' EMC effect  
{\it i.e.} on the modifications of the nuclear GPDs 
with respect to the free nucleon ones, normalized to their respective form
factors \cite{GuzStr,Scopetta,LiuTan2}.  
In \cite{LiuTan2}, in particular, it was shown that the role of 
partonic transverse degrees of freedom, accounted for by a careful consideration of 
nucleon off-shellness, 
is enhanced in the generalized EMC effect, with respect to the forward case. 
In addition, a number of interesting relationships were found by 
studying Mellin moments in nuclei:
the $A$-dependence for the $D$-term of GPDs was estimated
within a microscopic approach, and compared with the calculation of \cite{Polyakov} 
using a liquid drop model, 
and finally, a connection was made with the widely used approaches 
that relate the 
modifications of ``partonic'' parameters such as the string tension, or the confinement
radius, to density dependent effects in the nuclear medium (see {\it e.g.} \cite{Pirner}
and references therein). 
These results designate nuclear GPDs as a potentially important new tool 
to investigate nuclear hadronization and related phenomena, 
which are vital for interpreting current and future data 
from RHIC, LHC, HERMES, and Jefferson Lab.

The qualitatively new insight offered by nuclear GPDs calls both for a more detailed
study of the feasibility of experiments, and for more detailed studies
aimed at establishing a better connection 
between GPDs and observables. 
In this paper we concentrate on spin-0 nuclei, with the aim of disentangling 
those nuclear
effects that can be reconducted to the forward EMC effect \cite{EMC}. 
Studies of nuclei with different spin, such as the deuteron,   
are at variance with ours
since, due the more complex spin structure, 
they involve completely new functions with respect to the forward case
therefore making it less linear to investigate the nature of nuclear 
medium modifications.
\footnote{Our calculations however can be applied straightforwardly 
to the deuteron ${\cal H}_1$ function, or to channels where only one of
the deuteron functions dominates.}

The structure of the cross section for DVCS off a spin-0 nucleus 
involves the matrix elements in Eq.(\ref{T}), similarly to proton case, 
with few important changes that we 
describe below. In particular, the nuclear GPDs can be extracted from
similar types of asymmetries as for the proton case discussed above. 
Here we present a calculation for the nuclear beam spin asymmetry, $A_{LU}^{(A)}(\phi)$.
$A_{LU}^{(A)}$ has been recently measured using $D$, $Ne$, and $Kr$ targets 
at HERA \cite{Hasch}.  
Furthermore, a number of experiments are currently planned at HERA \cite{Hasch} and 
Jefferson Lab \cite{Jlab_prop}. It is therefore now mandatory 
to provide {\it quantitative} evaluations of the generalized parton distributions 
entering the definition of $A_{LU}^{(A)}$. This reads:
\begin{eqnarray}
A_{LU}^{(A)}  = \frac{d\sigma^\uparrow - d \sigma^\downarrow}
{d \sigma^\uparrow + d \sigma^\downarrow} 
 \approx   \frac{s_1^{\cal I}}{c_o^{BH}} \sin \phi.  
\label{ALU}
\end{eqnarray}
The full expressions for the coefficients $s_1^{\cal I}$ and $c_o^{BH}$  
can be found
in Appendix B of Ref.\cite{KirMul} and will not be repeated here.
One can extract from such expressions the dominant contribution to Eq.(\ref{ALU}) 
for a spin-0 nucleus, namely: 

\[ s_1^{\cal I} \propto \Im {\rm m} \, {\cal H}_A \, F_A(t), \]
\[c_o^{BH} \propto \left[ F_A(t) \right]^2, \]
$F_A(t)$ being the nuclear form factor. 
By adopting the notation of \cite{LiuTan2}
which extends the variables and functions defined in \cite{GolMar} to the nuclear case,  
one can write:   
\begin{equation}
\Im {\rm m} \, {\cal H}_A(X,\zeta,t) = -\pi \sum_q
e^2_q \left[ H_A^q(\zeta,\zeta,t) + H_A^{\bar{q}}(\zeta,\zeta,t) \right],
\label{ALU3}
\end{equation}
where we have introduced the nuclear GPDs, $H_A^{q (\bar{q})}$.

%
However, for a nuclear target 
two distinct classes of final states can occur, both in DVCS and BH:
{\it i)} the scattering can happen {\it coherently}, that
is the target nucleus recoils as a whole while emitting a photon with
momentum $q^\prime$ (Figs.\ref{fig1}a and \ref{fig1}c); 
{\it ii)} the nucleus undergoes breakup
-- or {\it incoherent} scattering -- the final photon being emitted from
a quasi-elastically scattered nucleon (Figs.\ref{fig1}b and \ref{fig1}d). 

The contribution from coherent scattering
(Figs.\ref{fig1}a, \ref{fig1}c) is given by 
a {\em product} of amplitudes:
\begin{eqnarray}
\label{I_coh}
& & {\mathcal{I}}_{coh}(\zeta,t) = {\cal K} \, {\cal F}_{DVCS}^A(\zeta,t) F_{BH}^A(t), 
\end{eqnarray}
with:
\begin{subequations}
\begin{eqnarray}
& & {\cal F}_{DVCS}^A(\zeta,t)  =  \int \frac{d^2 P_\perp dY}{2(2\pi)^3}
\mathcal{N} \, \rho^{A}(Y,P^2;\zeta,t) 
\nonumber \\
& & \times  {\cal F}^N_{DVCS} \left(\frac{\zeta}{Y},P^2; \frac{\zeta}{Y},t \right),
\label{DVCS_coh}
\\ 
& & F_{BH}^A(t) =  \int \frac{d^2 P_\perp dY}{2(2\pi)^3}
\mathcal{N} \, 
\rho^{A}(Y,P^2;\zeta,t)  F_1^N(t),
\label{BH_coh}
\end{eqnarray}
\end{subequations}
where the nucleon ($N$) amplitude, in the small $\zeta$ approximation, {\it i.e.} by 
disregarding $E^{q(\bar{q})}$, is given by:
\begin{eqnarray}
& {\cal F}^N_{DVCS} & \approx \sqrt{1-\zeta} \, \sum_q 
e_q^2 \left[ H^q(\zeta,\zeta,t) + H^{\bar{q}}(\zeta,\zeta,t) \right].
\label{FN_DVCS} 
\end{eqnarray}
Eq.(\ref{DVCS_coh}) was derived in \cite{LiuTan2}. 
In the forward limit -- $\zeta, t \rightarrow 0$ -- it yields 
the Deep Inelastic Scattering (DIS) nuclear structure function. 
Its ratio to the proton amplitude allows one to study the ``generalized'' EMC effect.
Eq.(\ref{BH_coh}) is by definition the nuclear form factor, namely 
$\displaystyle F^A(t) \equiv F^A_{BH}(t)$.

The kinematical variables in Eqs.(\ref{I_coh}-\ref{FN_DVCS}) are defined as: 
$Y=P^+/(P_A^+/A)$, the Light Cone (LC) nuclear 
momentum fraction taken by the struck nucleon (per nucleon),
$X/Y=k^+/P^+$, the LC momentum fraction taken by the struck quark,
$P^2$, the nucleon virtuality, 
$\zeta \equiv \Delta^+/(P_A^+/A)$, 
and $\zeta/Y =  \Delta^+/P^+$, the skewedness with respect to the 
nucleon.
Furthermore, ${\cal N} = (Y-\zeta/2)(\sqrt{Y(Y-\zeta)})$, is a normalization factor 
whose form derives from the normalization  
of the nucleon spinors in the off-forward case \cite{LiuTan2};
$\rho^{A}(Y,P^2; \zeta,t)$ is the off-forward
LC nuclear spectral function \cite{LiuTan2},  
accounting for all configurations 
of the final nuclear system (see \cite{CioLiu} and references
therein for definitions in the forward case). 
When the dependence on $P^2$ is disregarded, $\rho^{A}$ can be written within a 
non-relativistic approximation as: 
\[ \displaystyle \rho^{NR}_A(Y,\zeta,t)  =  2 \pi M \int_{P_{min}(Y,\overline{E})}^\infty 
d P P \Phi_A(P) 
\Phi_A^*(P^\prime) .\]
where $P$, and $P^\prime$ are the incoming and outgoing nucleons three-momenta, 
$\Phi_A$ is the nuclear wave functions in momentum space;
$\displaystyle \overline{E}$,
is the average separation energy characterizing the final nuclear system. 
Eq.(\ref{DVCS_coh}) then becomes a longitudinal convolution in the variable $Y$. 

For incoherent scattering, we consider the  
matrix elements for the nuclear DVCS and BH processes, respectively, 
with a break-up of the final nucleus. 
These can be read from Fig.\ref{fig1}b and Fig.\ref{fig1}d, respectively, as:
%
\begin{widetext}
\begin{eqnarray}
M^{A}_{inc-DVCS}  & = &  
 \left[\overline{U}(P',S')
\overline{\Gamma}(k',P)\frac{(\not\!{k'}+m)}{k'^2-m^2}\frac{(\not\!{k}+m)}
{k^2-m^2}{\Gamma}(k,P)U(P,S)\right] 
\left[\frac{(\not\!{P}+M)}{P^2-M^2}\Gamma_{A}(P,P_{A-1})
U_{A-1}(P_{A},S) \right], 
\\
M^{A}_{{inc-BH}^*} & = &
{\frac{1}{2P^{+}}}\left[{\overline{U}(P',S')}[{\gamma^{+}}F_{1}(t)+
\frac{i \sigma^{+\nu} \Delta_\nu }{2M} F_{2}(t)]
U(P,S)\right]
\left[\overline{U}_{A-1}(P_{A},S)\overline{\Gamma}_{A}(P,P_{A-1})
\frac{(\not\!{P}+M)}{P^2-M^2}\right]
\end{eqnarray}
\end{widetext}
where, similarly to Ref.\cite{LiuTan2}, the nucleon vertex is described by a 
quark-diquark model, $\Gamma$ being the vertex function,
$U(P,S)$, $U(P^\prime,S^\prime)$ the nucleon spinors; 
$\Gamma_A$ is the nuclear vertex function,  
and $U_{A-1}$ describes the $A-1$ system.

For incoherent scattering, the interference term contributing to Eq.(\ref{T}) 
is therefore given in the small $\zeta$ limit by a {\em convolution}:
\begin{eqnarray}
& & \mathcal{I}_{inc}(\zeta,t)  = {\cal K}
\int \frac{d^2 P_{\perp}{dY}}{2(2\pi)^3} \, \mathcal{N} \, \rho^{A}_0(Y,P^2) 
\nonumber \\ 
& \times & {\cal F}^{N}_{DVCS}\left(\frac{X}{Y},P^2;\frac{\zeta}{Y},t \right)
 F_1^N(t) \nonumber \\
& \equiv & {\cal F}^{A}_{DVCS, \, 0}(\zeta,t) F_1^N(t),
\label{I_inc}
\end{eqnarray}
where, within the nuclear impulse approximation illustrated in Fig.\ref{fig1}, 
no momentum is transfered to the nuclear vertex. Therefore, at variance with
coherent scattering, only the forward
LC nuclear spectral function, 
$\rho^{A}_0(Y,P^2) = \rho^{A}(Y,P^2;\zeta=0,t=0)$, enters the calculation.


Next, we consider the numerical impact of both the coherent and incoherent 
processes on the asymmetry, $A^{(A)}_{LU}$.
By including the contributions of both Eq.(\ref{I_coh}),
nd Eq.(\ref{I_inc}) in Eq.(\ref{ALU}), one has:
\begin{eqnarray}
A_{LU}^{(A)}& \propto & 
\frac{Z^2 \left[ {\cal F}_{DVCS}^A(\zeta,t) F_A(t) \right] 
+ Z \left[ {\cal F}_{DVCS, \, 0}^A(\zeta,t) F_1(t) \right]}
{Z^2 F_A^2(t) + Z F_1^2(t) } 
\nonumber \\
& \times & \sin \phi.
\label{ALU2}
\end{eqnarray}
where $Z$ is the number of protons, $F_1(t)$ is the proton form factor; 
in Eq.(\ref{ALU2}) the neutron contributions,  
$H_n$, $F_1^n$, were disregarded. Furthermore, similarly
to Ref.\cite{LiuTan2}, we consider only valence quarks contributions.
These are expected to dominate the asymmetries at the kinematics of 
\cite{CLAS,HERMES,Hasch}, namely at low $Q^2$ and intermediate values of Bjorken $x$, $x_{Bj}$.  
All form factors in Eq.(\ref{ALU2}) are normalized to unity and, correspondingly,
$H_d=(1/2)H_u$.
  
By setting: 
\[ {\cal I}_{coh}^A = \left[ {\cal F}_{DVCS}^A(\zeta,t) F_A(t) \right] , \; \; 
{\cal I}_{incoh}^A = \left[ {\cal F}_{DVCS, \, 0}^A(\zeta,t) F_1(t) \right], \] 
the ratio of the nuclear to proton asymmetry at leading order can be written as:
\begin{widetext}
\begin{eqnarray}
R_{LU}^{(A)}(\zeta,t) =  
\frac{Z^2  {\cal I}_{coh}^A
+ Z {\cal I}_{incoh}^A } 
{ {\cal F}_{DVCS}^p(\zeta,t) F_1(t)} \times 
\frac{F_1^2(t) }{Z^2 F_A^2(t) + Z F_1^2(t) } 
\label{A_rat}
\end{eqnarray}
\end{widetext}
Notice that in the absence of incoherent scattering, and 
in the small $\zeta$ limit, $R_{LU}^{(A)}$  
becomes the off-forward EMC effect ratio calculated in \cite{LiuTan2}:
\begin{eqnarray}
R_A (\zeta,t) = \frac{H^A(\zeta,t)/F^A(t)}{H^N(\zeta,t)/F^N(t)}.  
\end{eqnarray}

\begin{figure}
\includegraphics[width=6.5cm]{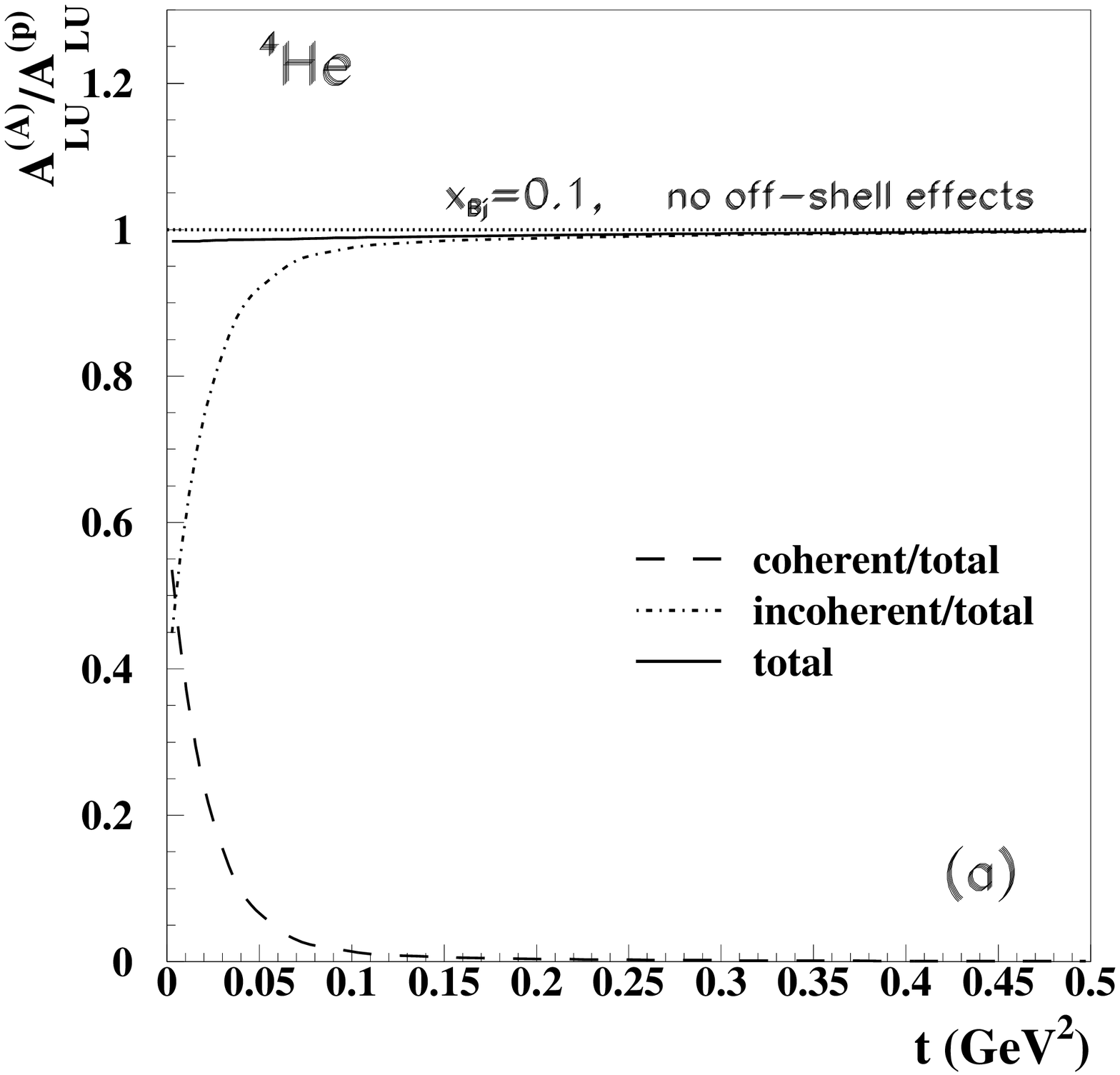}
\includegraphics[width=6.5cm]{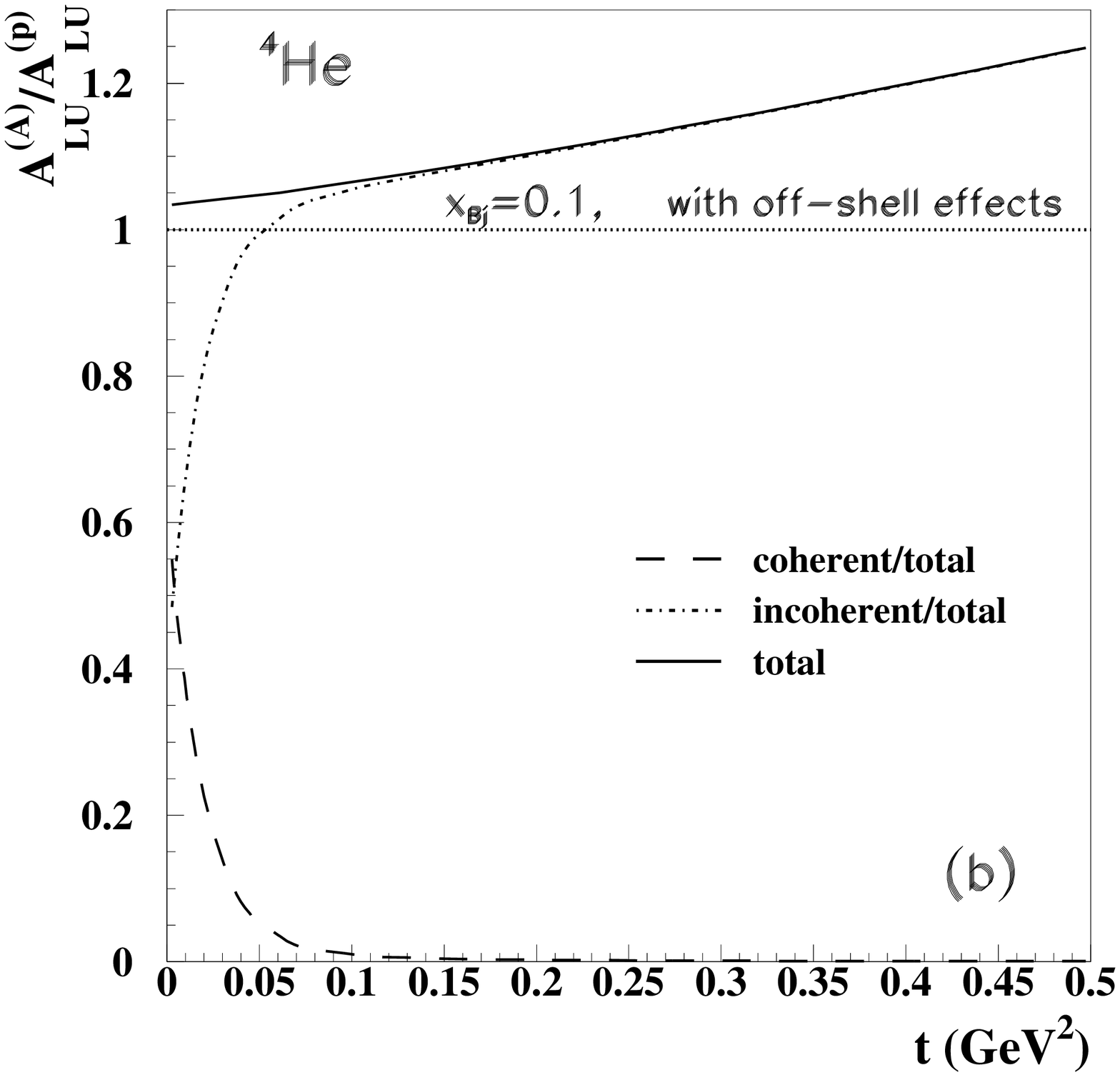}
\caption{Ratios $R_{LU}^{(A)}$, Eq.(\ref{A_rat}), (full line),  
$R_{LU}^{coh \, (A)}$, Eq.(\ref{arc}), (dashed line)
and $R_{LU}^{incoh \, (A)}$,  Eq.(\ref{ari}), (dot-dashed line), evaluated at $x_{B j}=0.1$
and $0 < t < 0.5$ GeV$^2$. In {\bf (a)} 
nuclear effects were taken into account using a longitudinal convolution formula,
{\it i.e.} $X$($\zeta$)-rescaling; in {\bf (b)} nucleon off-shell effects were taken 
into account (see Ref.\cite{LiuTan2} for more details).  
In both cases, 
incoherent scattering dominates the asymmetry for $t \gtrsim 0.05$ GeV$^2$. An enhancement 
at $t \approx 0.1$ GeV$^2$, corresponding to the kinematics of Ref.\cite{Hasch} was found,
consistent with the preliminary data.}
\label{fig2}
\end{figure}

\begin{figure}
\includegraphics[width=7.5cm]{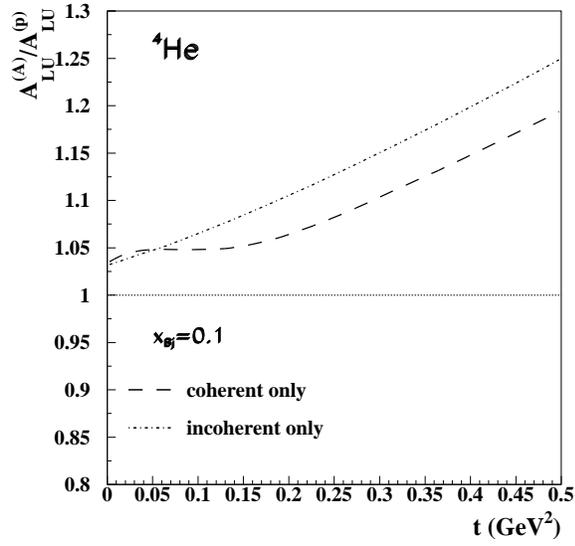}
\caption{Ratios $R_{LU}^{(A)}$, Eq.(\ref{A_rat}), calculated including only coherent 
scattering terms in both the DVCS and BH contributions to the asymmetry (dashed line), 
and including only the incoherent terms (dot-dashed line). The nuclear model including off-shell
effects was used in the calculations. Notice that the coherent contributions
correctly reproduce the off-forward EMC effect calculations of Ref.\cite{LiuTan2}. 
Same kinematics as in Fig.2.}
\label{fig3}
\end{figure}

The behavior of $R_{LU}^{(A)}(\zeta,t)$ is shown in Figs.\ref{fig2} and 
\ref{fig3}. In Fig.\ref{fig2} we plot the following quantities: 
{\it i)} $R_{LU}^{(A)}(\zeta,t)$, as given in Eq.(\ref{A_rat}); 
{\it ii)} the DVCS coherent contribution to the ratio:
\begin{eqnarray}
R_{LU}^{coh \, (A)}(\zeta,t) =  
\frac{Z^2  {\cal I}_{coh}^A} 
{ {\cal F}_{DVCS}^p(\zeta,t) F_1(t)} \times 
\frac{F_1^2(t) }{Z^2 F_A^2(t) + Z F_1^2(t) };  
\label{arc}
\end{eqnarray}
{\it iii)} the DVCS incoherent contribution to the ratio:
\begin{eqnarray}
R_{LU}^{incoh \, (A)}(\zeta,t) =  
\frac{Z  {\cal I}_{incoh}^A} 
{ {\cal F}_{DVCS}^p(\zeta,t) F_1(t)} \times 
\frac{F_1^2(t) }{Z^2 F_A^2(t) + Z F_1^2(t) },  
\label{ari}
\end{eqnarray}
All calculations were performed for the $^4He$ nucleus, as a prototype of spin-0 complex nuclei
\footnote{Calculations for larger nuclei are available upon request at the e-mail addresses 
listed within.}
using the microscopic nuclear model
described in Ref.\cite{LiuTan2}. Results were summarized in two panels, Fig.\ref{fig2}a and
Fig.\ref{fig2}b, in order to better illustrate the role played by particles'
off-shellness within our microscopic nuclear model. As explained in detail in Ref.\cite{LiuTan2} 
the account of nucleons' off-shellness translates into a different, $A$-dependent relation
between the struck quark's transverse degrees of freedom in a proton and in a nucleus, 
respectively, which is emphasized in off-forward observables such as GPDs. 
Off-shell effects are in fact quite noticeable, as seen by comparing 
Fig.\ref{fig2}a, where they were disregarded, and 
Fig.\ref{fig2}b. Moreover, both panels show 
Eqs.(\ref{A_rat},\ref{arc},\ref{ari}) at $x_{Bj} =0.1$ and in the range: $0 < t < 0.5$ GeV$^2$. 
Clearly, incoherent DVCS dominates the asymmetries at $t \gtrsim 0.05$ GeV$^2$, that is in the
range of recent HERMES data \cite{Hasch}. Also, our microscopic approach predicts an enhancement
which is consistent with the HERMES data. Fig.\ref{fig3} shows the ``coherent'' and ``incoherent''
contributions separately, obtained by including also 
separately the contributions in the BH term in Eqs.(\ref{arc}) and (\ref{ari}).
These contributions could be observed by detecting one of the outgoing
nuclear constituents, as planned in future experiments \cite{Hasch}. 
Our calculation points out that the nuclear effects are expected to similar, in both cases,
since they originate from similar modifications of the deep inelastic structure of the nucleus,
encoded in our model in Eq.(\ref{DVCS_coh}) for coherent, and Eq.(\ref{I_inc}) for incoherent 
scattering (the two equations differ only in the nuclear $t$ dependence, that does not affect
the deep inelastic structure). 

In conclusion, we presented for the first time quantitative evaluations of the coherent
and incoherent contributions to the ratio of the nuclear beam spin asymmetry over the
proton one. Although incoherent scattering dominates current experiments, interesting
information can still be extracted relating the off-forward EMC effect to the previously
measured, and still for many aspect puzzling, forward EMC effect. The difference between 
the coherent and incoherent cases can in fact be traced to the replacement of the 
off-forward LC nuclear spectral function with the forward one in the coherent case. This
affects the $t$ dependence of the nuclear wave functions, and has little impact on the
modifications of the deep inelastic structure of a bound nucleon.      


\end{document}